\input epsf.tex
\documentstyle[prl,aps]{revtex}
\input epsf.tex
\def\psid{{\psi^\dagger}}
\def\Lag{{\cal L}}
\def\sigmav{{\vec \sigma\,}}
\def\expectop#1#2#3{{ \langle {#1} | {#2} | {#3} \rangle }}
\def\eqn{\begin{equation}}
\def\ee{\end{equation}}
\def\be{\begin{equation}}
\def\frac#1#2{{#1 \over #2}}
\def\endeqn{\end{equation}}
\def\ba{\begin{eqnarray}}
\def\ea{\end{eqnarray}}
\begin{document}
\draft
\title
{ A new ${\cal O}(\alpha)$ correction
to the decay rate of pionium 
}
\author{Patrick Labelle \thanks{e-mail: labelle@hep.physics.mcgill.ca} }
\address{ Physics Department ,
McGill  University, Montreal , Canada  H3A 2T8}
\author{and Kirk Buckley}
\address{Physics Department, Bishops' University, Lennoxville, Canada}
\date{\today}
\maketitle
\begin{abstract}
Recently, much work has been devoted to the calculation of ${\cal O}
(\alpha)$ corrections to the decay rate of pionium, the $\pi^+
\pi^-$ bound state. In previous calculations,
nonrelativistic QED  corrections were neglected since they 
start at order $\alpha^2$ in hydrogen and positronium.
In this note, we point out that there is one correction which
is actually of order $\alpha$ times a function of 
the ratio
$\mu_r \alpha/m_e$,
where $\mu_r$ is the reduced mass of the system. When $\mu_r 
\alpha  \ll m_e$, this function can be Taylor expanded and
leads to higher order corrections.  When $\mu_r \alpha
\approx m_e$, as is the case in pionium, the function is of order
one and the correction is of order $\alpha$. We use 
an effective field theory approach to calculate this correction and
 find it equal to $0.4298 \, \alpha
\Gamma_0$. We also calculate
 the corresponding contribution to the dimuonium
($\mu^+ \mu^-$ bound state) decay rate and find 
 a result  in agreement
with  calculations by Jentschura  {\it et al}.  
\end{abstract}

\vspace{5ex}

\section{Introduction }
\label{sec:INTRO}

The decay rate of pionium ($\pi^+ \pi^-$ bound state) has generated
much interest recently, because of the possibility of extracting from
a future experiment at CERN \cite{cern1} \cite{cern2}  precise values of the
pion scattering lengths  which can, in turn, be related
to  the quark condensate
parameter, a fundamental ingredient of ChPT. In
order to do so, however, all corrections of order $\alpha$
times the lowest order decay rate must be evaluated. These fall in
several categories and have been the subject of several recent papers
\cite{jal} \cite{ruset} \cite{kn}.

    One can distinguish between pure
QED corrections 
and corrections which arise from the interplay of strong
and QED effects. The former are associated with an expansion in
$\alpha$ whereas the latter contains an expansion
in  powers of $\alpha$  and other 
ChPT coefficients, such as scattering lengths,
 the pion mass difference,
etc.

 The pure QED corrections can be subdivided in
relativistic and nonrelativistic contributions. This separation of
scales is used, in one form or another, in all nonrelativistic bound
state calculations but is the most transparent  in the effective field
theory (eft)  formalism. The eft appropriate to the study of
nonrelativistic QED bound states, NRQED,
 was developed by Caswell and Lepage
more than ten years ago \cite{NRQED}. NRQED has since been
used to calculate high order corrections to the hyperfine
splitting of both muonium \cite{nio} and 
positronium \cite{hfs6}. Using NRQED, it is easy
to convince oneself that in hydrogen
and positronium, the  ${\cal O} (\alpha)$  QED corrections 
are purely relativistic in nature whereas the
nonrelativistic corrections start at order $\alpha^2$. 
This argument was used
in \cite{jal} to dismiss the nonrelativistic corrections\footnote
{ In that reference, the nonrelativistic corrections are  
referred to as ``pure QED corrections in the channel $\pi^+ \pi^- -
\pi^+ \pi^-$".}. 

In this note, we want to point out that the vacuum
polarization correction to the Coulomb interaction
(referred to as the ``VPC" correction in the
following) is actually of order $\alpha$
times a function of the dimensionless ratio $
\mu_r \alpha / m_e$, where $\mu_r$ is the
reduced mass of the bound state. In hydrogen
and positronium, this ratio is very small ($\approx \alpha$
and $\alpha/2$, respectively) and the function can be Taylor
expanded, leading to a correction of higher order.
In pionium, however, the ratio is equal to $0.997$
and the function turns out to be of order one. This
means that the VPC correction represents an
${\cal O} (\alpha)$ correction to the lowest order rate. 
We want to emphasize that the observation that the VPC
correction is enhanced in atoms with large reduced mass
 is not new;
it is, for example, discussed in the Landau and Lifshitz 
textbook on quantum electrodynamics \cite{LL}, in the context
of energy levels.

Another system in which this type of correction is large is dimuonium,
the $\mu^+ \mu^-$ bound state. In that system,
 the ratio $\mu_r \alpha /m_e$ is about $0.754$ and the 
VPC correction is also  of order $\alpha$ times the lowest 
order decay rate. This contribution was
evaluated, among several 
other corrections, in a recent paper by
Jentschura {\it et al} \cite{karshen}. 
In this note, we use effective field theories to
 calculate the VPC contributions to both
dimuonium and pionium. Our dimuonium result,
$0.337 \alpha \Gamma_0$, agrees with  \cite{karshen}
 Our result for pionium is
equal to $0.430 \, \alpha \Gamma_0$,  which is as large as the other 
corrections evaluated in
\cite{jal} \cite{ruset}.

We first consider the VPC corrections to the decay rate 
of dimuonium. For the present calculation, the only relevant interactions
 of the
NRQED Lagrangian are 
\ba
\Lag_{fermions}~& =  &\psid\bigg\{ i \partial_{t} +
\frac{\vec D^2}{2m_\mu} -e A_0) \biggr\} \psi 
 ~+~ {\rm same~terms~with~} \psi \rightarrow \chi 
\cr &-& 
 c_4 ~  \psi^\dagger \sigmav  \sigma_2
\chi^*\,  \cdot\,  \chi^T
 \sigma_2 \sigmav
 \psi
-~c_{5} ~\psi^\dagger  \sigma_2 \chi^*\, \cdot \,
  \chi^T   \sigma_2 \psi
\ea
where $\psi$ refers
 to the muon field, $\chi$ refers to the antimuon
field and $m_\mu$ is the muon mass. Our
sign convention for the coefficients $c_4$ and
$c_5$ differs from \cite{cliff} in order 
to avoid minus signs in the Feynman rules.
 To the order of interest, we can approximate
the gauge derivative $\vec D $ by $ \vec p$ since the
correction due to  transverse photons are suppressed
(see \cite{MQED}). The Feynman rules of the relevant
interactions are presented in Fig.[1].
We  work in Coulomb gauge in which the Coulomb photon
propagator is simply given by $-1/ \vec k^2$. The vacuum correction to
the Coulomb propagator is also illustrated in Fig.[1]. We consider
only 
the VPC because it is the only interaction which contributes
at order $\alpha $ times a function of $\mu_r \alpha / m_e$.
  The vacuum polarization correction
to the transverse propagator can be neglected because the vertices
connecting 
transverse photons to fermions are suppressed by at least one
power of the muon mass, which translates
 into additional factors of $\alpha$
in the bound state calculation (the interested reader is referred to
\cite{MQED} for more details on NRQED power counting rules).  

 The four-fermion operator
proportional to $c_4$ leads to the decay rate of orthodimuonium
(dimuonium with no angular momentum and a total spin equal to 1)
whereas the other four-Fermi operator contributes to the decay of
paradimuonium ($L=0, S=1$). In a nonrelativistic system, the Coulomb
interaction is nonperturbative and must be summed up to all orders,
which is equivalent to solving the Schr\"odinger equation. We will focus
on the decay rate of the ground state ($n=1$), for which the
wavefunction is given by
\be
\Psi(\vec p) = { 8 \pi^{1/2} \gamma^{5/2} \over ( \vec p^2 + \gamma^2) }
 \otimes  \xi_{\mu^+} \otimes \xi_{\mu^-} 
\ee
where  the $\xi$'s  are   the
two component Pauli
spinors of the muon  and antimuon and $\gamma \equiv \mu_r \alpha$  
represents the typical bound state momentum (the ground state energy
is given by $-\gamma^2/(2 \mu_r)$).

 In NRQED,  bound state properties are
 computed in two steps. First, one
fixes the NQRED coefficients by matching 
 QED and NRQED scattering amplitudes, and then 
one computes the bound state energy by applying time ordered
perturbation theory to the NRQED vertices (with the Coulomb interaction
defining the unperturbed problem). The decay rate is given in terms of
the imaginary part of the energy by 
$\Gamma= $ -2 Im (E) .

To illustrate the use of NRQED, we now focus on the decay rate of
orthodimuonium.
To simplify the following discussion, we define the operator
\be
{\cal O}(\,^3S_1) \equiv {\rm Im}  (c_4)   
\vec S^2\ee
where the spectroscopic
notation $\,^{2S+1}L_J$
is used to indicate the quantum numbers of 
a state annihilated by this operator.
 In terms
of this operator, the decay rate is given by 
\be
\Gamma (1^3S_1) = - 2~ \expectop{\Psi}{{\cal O}(\,^3S_1)}
{\Psi} 
=  -4 \vert \Psi(\vec r = 0) \vert^2
  {\rm Im} (c_4)  ~=~- {m_{\mu}^3 \alpha^3
\over 2 \pi} {\rm Im} (c_4) \label{decay}
\ee
where a factor of 2 comes from the spin average of
 $\vec S^2$ and $\vert \Psi(\vec r = 0)\vert^2 =
\gamma^3/ \pi$.
 The imaginary part of the
 coefficient $c_4$ contains an infinite expansion
in powers of $\alpha$, the lowest order contribution being 
found by performing the matching
illustrated in Fig.[2]. All diagrams are computed exactly at threshold
(no external momenta), in the center of mass frame of the 
particles, and
a
nonrelativistic normalization is used for the QED diagram ({\it
i.e.} a factor of $1/\sqrt{2E} = 1/\sqrt{2m}$ is provided for each
external particle). Also, we match the amplitude matrix ${\cal M}$ and
not $-i {\cal M}$. The superscript 1 on $c_4$
indicates that this is a contribution coming from a one-loop QED
diagram. The NRQED diagram is simply $2  \,
{\rm Im}  (c_4^{(1)}) $
 whereas 
the imaginary part of the
QED diagram   is $-2 \pi \alpha^2/(3 m_\mu^2)$. 
Solving for the NRQED coefficient, we get  
\be
{\rm Im} (c_4^{(1)})=-{\pi \alpha^2 \over 3 m_\mu^2 } 
\label{yaya}
\ee
which, after putting in Eq.(\ref{decay}) gives the  lowest order rate, 
\be
\Gamma_0(1^3S_1)= {m_{\mu} \alpha^5 \over 6}.
\ee
We now turn to  the contribution from the vacuum polarization
correction to the Coulomb photon. The first thing to do is to revise
the matching of ${\rm Im}( c_4)$ beyond one-loop. At 
first, this might seem unnecessary since we are interested
in ${\cal O} ( \alpha \Gamma_0)$ corrections
only, but we must be careful about factors of $m_\mu/m_e$ which
might compensate for factors of $\alpha$. Indeed, this situation
occurs in the matching  
 illustrated in Fig.[3]. The result for 
the imaginary part of the  QED diagram can be found from \cite{hoang}:
\be
 2\times({\alpha \over \pi})^2 \biggl({9 \zeta(2) \over 16} 
{m_{\mu} 
\over m_e} + {4 \over 3} \ln({m_{\mu} \over m_e}) + {22 \over 
36} +{\cal O} ({m_e \over m_{\mu}}) \biggr) \bigl\{ 
-{2  \pi \over 3}  {\alpha^2 \over m_\mu^2}\bigr\}
\ee
where the term in curly brackets
 is simply the lowest order result and the overall
factor of 2 takes into account the two
permutations of the
diagram.  The  term proportional
to $m_\mu / m_e$
represents a correction  equal to 
  $ 0.283 \, \alpha
$ and should therefore be treated as an ${\cal O} (\alpha)$
correction instead of an $\alpha^2$ correction. To the
precision of interest in this work, we can neglect the 
log term, which comes from the running of the coupling
constant, and all the other terms not enhanced by a 
factor of $m_\mu / m_e$.
The calculation of the NRQED one-loop scattering diagram 
in Fig.[3] is 
straightforward:
\be
\bigl\{- {2 \pi \over 3} {\alpha^2 \over m_\mu} \bigr\}
~\int {d^3 k \over (2 \pi)^3} {-2 \mu_r \over
\vec k^2} ~\int dv~{-4 \alpha^2 v^2(1-v^2/3) \over
\vec k^2 (1 - v^2) + 4 m_e^2} ~=~
{3 \over 16} {m_{\mu} \over m_e} \alpha^2 \bigl\{-{2 \pi \over 3}
{\alpha^2 
\over m_\mu^2} \bigr\}
\ee
which is seen to cancel exactly the $m_{\mu}/m_e$ term in the QED
diagram. This is not a coincidence. The $m_{\mu}/m_e$ term is
associated to the low energy behavior of the QED diagram,
and since NRQED is designed to reproduce QED in the nonrelativistic
limit, this term had to be present in the NRQED diagram. This
will hold for higher order loop diagrams as well. 
The end result is that the coefficient ${\rm Im} (c_4)$
 does {\it not} receive any 
correction of order $\alpha (\alpha m_{\mu}/m_e)^n
\approx \alpha $
from the matching to QED. Therefore,
we can simply use 
 the lowest order coefficient $
{\rm Im} (c_4^{(1)})$  given in Eq.(\ref{yaya}) in the bound state
calculation.


The bound state contribution of the VPC interaction is found by 
applying   
 second order
perturbation theory:
\be 
\delta \Gamma_{vpc}(\,^3S_1) = 
2~ \langle \, n \, | {\cal O} (\,^3 S_1) \sum_{m \ne n} 
\hspace{-6mm}\int
 { | \,m\, \rangle \, \langle \, m \, | \over E_n - E_m }
\, V_{vpc} \, | \, n \, \rangle   
\ee
where $V_{vpc}$ is the potential corresponding to the vacuum
polarization correction to the  Coulomb interaction:
\be
V_{vpc} ~=~  \int_0^1 dv { - 4 \alpha^2
 v^2 ( 1 - v^2/3) \over \vec
k^2(1-v^2) + 4 m_e^2} 
.
\ee
For the sum over intermediate states, we must include the full Coulomb
Green's function $G(\vec p, \vec q)$ in terms of which the previous
expression can be written as
\ba
\delta \Gamma_{vpc} (\,^3S_1) =
 -8 \times &&  \int  {d^3 p' \over (2 \pi)^3}
{8 \pi^{1/2} \gamma^{5/2} \over (\vec p'^2 + \gamma^2)}
{ - \pi \alpha^2 \over 3 m_\mu^2} 
 \cr && \int {d^3p \,  d^3q \,  d^3 k \over (2 \pi)^9} ~G(\vec p, \vec
q) \cr && ~ \biggl(  \int_0^1 dv { - 4 \alpha^2
 v^2 ( 1 - v^2/3) \over \vec
k^2(1-v^2) + 4 m_e^2} \biggr) { 8 \pi^{1/2} \gamma^{5/2} \over (\vec p -
\vec k)^2 + \gamma^2 } 
\ea
\be
=2  \int {d^3p \,  d^3q \,  d^3 k \over (2 \pi)^9} ~G(\vec p, \vec
q) ~ \biggl(  \int_0^1 dv { - 4 \alpha^2
 v^2 ( 1 - v^2/3) \over \vec
k^2(1-v^2) + 4 m_e^2} \biggr) { 8 \pi \gamma \over (\vec p -
\vec k)^2 + \gamma^2 } ~ \Gamma_0
\label{decay2}
\ee
where,
 in the
 first expression, a factor of -2 comes from
the relation  $\Gamma = - 2 $ Im(E), a factor
of 2 comes from the spin average, and another one comes from
the two sides on which the interaction can take place.
The second expression gives the correction in terms of the lowest order rate,
$\Gamma_0$. 
For the Coulomb Green's function, we use the following  expression 
derived
by Schwinger \cite{schwinger}:
\be
G(\vec p, \vec q) ~=~-{2 \mu_r \over \vec p^2 + \gamma^2} (2 \pi)^3
\delta^3(\vec q)  - { 16 \pi \mu_r^2 \alpha \over (\vec p^2 + \gamma^2)
(\vec p - \vec q)^2 (\vec q^2 + \gamma^2) } - {64 \pi \over \alpha
\gamma^4} R(\vec p, \vec q)
\ee
where the first term corresponds to no Coulomb interaction
 in the intermediate
state, the second term corresponds to one Coulomb line and the third
term corresponds to the the  exchange of two or more Coulomb photons.
We will refer to these three terms as the $0-C$, $1-C$ and $R$
contributions, respectively. A graphical representation of
Eq.(\ref{decay2}) is presented in Fig.[4]. 
An explicit expression for $R(\vec p, \vec q)$ can be found
 in \cite{schwinger}.
All we need for the present work is the following integral \cite{Lepage}
\be
\int (d^3p d^3 q) f(p) R(\vec p, \vec q)
~=~-  4 \pi^3 \gamma^7 ~\int_0^\infty\, dp~ {p^2 \, f(p)
\over ( \vec p^2 + \gamma^2)^2 } \biggl(\ln (2) - {5
\over 2} + {\gamma \over p}
{\rm Arctan} ({p \over \gamma}) - {1 \over 2}
 \ln (1 + {\vec
p^2 \over \gamma^2} ) + {4 \gamma^2 \over \vec p^2 + \gamma^2}
\biggr) \label{rr}
\ee
where $f(p)$ is an arbitary function of the magnitude of the
three-momentum $\vec p$.

We now  calculate the correction to the decay rate of
orthodimuonium. The no Coulomb correction to Eq.(\ref{decay2}) 
can be integrated analytically and is found to be
\be
\Delta E_{0-C}~=~{2 \over \pi^3} \,I_0(r)~ \alpha \Gamma_0 
\label{NoC}
\ee
where $I_0$ is a function of the dimensionless ratio $r\equiv 
\gamma / m_e 
= \mu_r \alpha / m_e$ which is $0.754$ in dimuonium.
For the sake of completeness, we give $I_0(r)$ for arbitrary values 
of
r:
\ba
I_0 &=& \pi^2 \biggl( -{5\over 9} -{1 \over 3 r^2} +
{\pi \over r} \bigl( {1  \over 4} + {1 \over 6 r^2} \bigr) + 
{\bigl( -2 r + 1/r + 1/r^3\bigr) \over 3 {\sqrt{1-r^2}} }
\biggl\{ {\rm ArcTan} \bigl( {r \over {\sqrt{1-r^2}}} \bigr)
 - {\pi \over 2} \biggr\} \biggr)  ~~~(r <1)
\cr &=& {\pi^2 \over 36 } \bigl(-32+15 \pi \bigr) ~~~(  r=1)
\cr &=& \pi^2 \biggl(-{5\over 9} -{1 \over 3 r^2} +
{\pi \over r} \bigl({1 \over 4} + {1 \over 6 r^2} \bigr) -
{\bigl(-2 r + 1/r + 1/r^3\bigr) \over 6 {\sqrt{r^2-1}} }
\biggl\{{\rm ln}  \bigl({ {\sqrt{r^2-1}}+r  \over 
r-{\sqrt{r^2-1}}} \bigr) \biggr\} \biggr) ~~~(  r > 1) 
.
\ea
Using Eq.(\ref{NoC}) with $r=0.754$, the no Coulomb 
contribution to the  dimuonium decay rate 
is found to be \be
 \Delta E_{0-C}({\rm dimuonium})~=~0.216 \alpha \Gamma_0~~\label{nocdm}
.
\ee
We have reduced the one Coulomb and R term contributions 
of Eq.(\ref{decay2}) 
to two dimensional integrals (using the identity (\ref{rr}))
which
were evaluated 
using the adaptive Monte-Carlo integration
routine Vegas \cite{vegas}.
 The results are respectively
\be
\Delta E_{1-C} ({\rm dimuonium}) ~=~ 0.08674 \, \alpha \Gamma_0 
\label{ocdm}
\ee
\be
 \Delta E_{R } ({\rm dimuonium}) ~=~
 0.0344 \, \alpha \Gamma_0
 \label{rtdm}
\ee
leading to a total equal to
 (sum of Eqs.(\ref{NoC}), (\ref{ocdm}) and (\ref{rtdm}))
 $0.337 \alpha \Gamma_0$. This result agrees with 
Ref.\cite{karshen} which gives $(1.06 / \pi) \alpha \Gamma_0 = 0.337 
\alpha \Gamma_0$. Notice that the lowest order decay rate of
paradimuonium is also 
given in terms of a  four-Fermi interaction
(the interaction proportional to $c_5$ in the Lagrangian)
so that  the VPC 
correction
is also $0.337 \, \alpha \Gamma_0$
where $\Gamma_0$ is now the lowest order paradimuonium
decay rate. 


The calculation of the VPC correction to the pionium
decay rate is very similar to the dimuonium calculation.
This is the most apparent by using   a
nonrelativistic field theory reproducing
ChPT at low energy, which we will refer to as ``NRChPT". 
 The only  NRChPT interactions relevant to the present calculation
are the Coulomb interaction (for the charged pions) and the 
four-pion vertex of the form $\pi^+ \pi^- \pi^0 \pi^0$
 (a more detailed presentation
of NRChPT and calculations of the other 
${\cal O} (\alpha)$ corrections will be presented elsewhere).

The matching of the four pions interaction is presented in
Fig.[5].
At threshold, the ChPT amplitude for the $\pi^+ \pi^- \pi^0 \pi^0$
vertex is equal to $32 \pi(a_0^0 - a_0^2)/3$, where $a_0^i$ 
is the scattering amplitude in the isospin channel $i$. Using a
nonrelativistic normalization for the external states, we
divide by a factor 
\be ({\sqrt{2E_{\pi^0}}})^2 {\sqrt{2 E_{\pi^+}}} {\sqrt{2 E_{
\pi^-}}}
\approx 4 m_{\pi^0} m_{\pi^+} \approx 4 m_{\pi^+}^2
\ee
where we have set $\Delta m_\pi \equiv
m_{\pi^+} - m_{\pi^0} = 0$. As we will
see below, this is possible because we are interested
in the decay reate to leading order in $\Delta m_\pi$ (the
${\cal O} (\Delta m_\pi)$ corrections will be treated
in a future publication). The end result of the matching is that 
the coefficient of the  NRChPT four-pion interaction 
is
\be
C_{4-pions}~=~ {8 \pi  \over 3} {(a_0^0-a_0^2) \over
 m_{\pi^+}^2 } .
\ee

For the coulomb photon and the VPC interaction, we
use the same Feynman rules as in the dimuonium calculation. 
As before, the Coulomb interaction must be summed up to
all orders, which means that the $\pi^+ \pi^-$
state will be described by the
same ground state Schr\"odinger wavefunction
as before.
The lowest order rate is trivially computed by 
considering the NRChPT interaction
sandwiched  between 
wavefunctions:
\be
\Delta {\rm E} ~=~{1 \over 2} \int{d^3 p \over (2 \pi)^3 }
\Psi(\vec p) ~C_{4 pions} \int {d^3q \over
(2 \pi)^3 } ~G_{\pi^0 \pi^0}(\vec q^2)~C_{4 pions}
\int {d^3 p' \over (2 \pi)^3}  \Psi( \vec p')
\label{lo}
\ee
where the overall factor of $1/2$ is a symmetry factor 
and $G_{\pi^0 \pi^0}(\vec q^2)$ is the free propagator
of the $\pi^0 \pi^0$ pair, given by
\be
G_{\pi^0 \pi^0}(\vec q^2) = {1 \over E_0 - E_{intermediate} }
\approx {1 \over 2 m_{\pi^+} - 2 m_{\pi^0} - \vec q^2/m_{\pi^0}
+ i \epsilon} \label{green}
\ee
where we have neglected the binding energy in $E_0$ (the
full expression is $E_0 = 2 m_{\pi^+} ( 1-  \alpha^2/8)$)
since it leads to a correction of order $\alpha^2 \Gamma_0$.
The difference $2 m_{\pi^+} - 2 m_{\pi^0}= 2 \Delta m_{\pi}$
appearing in Eq.(\ref{green})
cannot be set to zero since it leads to the imaginary part of 
the integral and, therefore, to the first order decay rate.
A trivial integration gives, for the integral Eq.(\ref{lo})
\be
\Delta E ~=~
- \vert  \Psi(0)\vert^2
{16 m_{\pi^0}  \over 9  }{ (a_0^0 - a_0^2)^2 \over
m_{\pi^+}^4 }
 \bigl(\Lambda + i {\pi
\over 2} {\sqrt{m_{\pi^0}}}
{\sqrt{ 2 \Delta m_{\pi}}} \bigr) \label{res}
\ee
where $\Lambda$ is an ultraviolet cutoff on the $\vert \vec q
\vert $
integration which would be canceled by a counterterm
in a calculation of the bound state energy \cite{thesisp}. Since we
are only interested in the decay rate, we need only to concentrate
on the imaginary part of the energy. Working
in first order in $\Delta m_{\pi}$, Eq.(\ref{res}) finally
leads to a decay rate equal to the well-known result
\be
\Gamma_0 = - 2 {\rm Im} (E) = {16 \pi \over
9} {\sqrt{2 \Delta m_{\pi} \over m_{\pi^+} }} {(a_0^0 - a_0^2)^2
\over m_{\pi^+}^2} ~\vert \Psi(0) \vert^2 .
\ee
Turning now to the VPC correction, we have to carry
out the integral given in Eq.(\ref{decay2}), using the reduced mass of
pionium, $\mu_r = m_{\pi^+}/2$. The no-Coulomb term is clearly 
given by  Eq.(\ref{NoC}), with $r=m_{\pi^+} \alpha /(2 m_e)
\approx 0.997$. We find a result equal to
\be
\Delta E_{0-C} ({\rm pionium}) ~=~0.26677 
\, \alpha \Gamma_0. \label{ncp}
\ee
We have also computed the one Coulomb and R term
contributions to the pionium decay rate
using VEGAS  and found 
\be
\Delta E_{1-C} ({\rm 
pionium})~=~ 0.1193 \, \alpha \Gamma_0 ~
 \label{ocp}
\ee
\be \Delta E_{R}
({\rm pionium}) ~=~ 0.04372 \,\alpha \Gamma_0 
. \label{rtp}
\ee
Our final result is the sum of Eqs.(\ref{ncp}), (\ref{ocp}) and
(\ref{rtp}):
\be
\Delta E_{vpc} ({\rm pionium})~=~0.4298 \, \alpha \Gamma_0
\ee
which is of the same order of magnitude as the other 
corrections calculated in \cite{jal} and \cite{ruset}.

\acknowledgments

P.L. Is indebted to several colleagues for very useful
exchanges. He  would like to first  thank J\"urg Gasser 
for introducing him to the problem of the pionium
decay rate calculation and for several hours of fruitful
conversations.  He would
also like to thank S. Karshenbo\v{\i}m for
very useful exchanges  on the dimuonium
calculation of \cite{karshen}, and Andr\'e Hoang for enlightening
comments on the $m_{\mu}/m_e$ terms in the QED diagrams.
Finally, he is indebted to 
A. Rusetsky for 
clarifying  discussions on \cite{ruset}.

\begin{figure}
\centerline{\epsfxsize 5.0 truein
\epsfbox{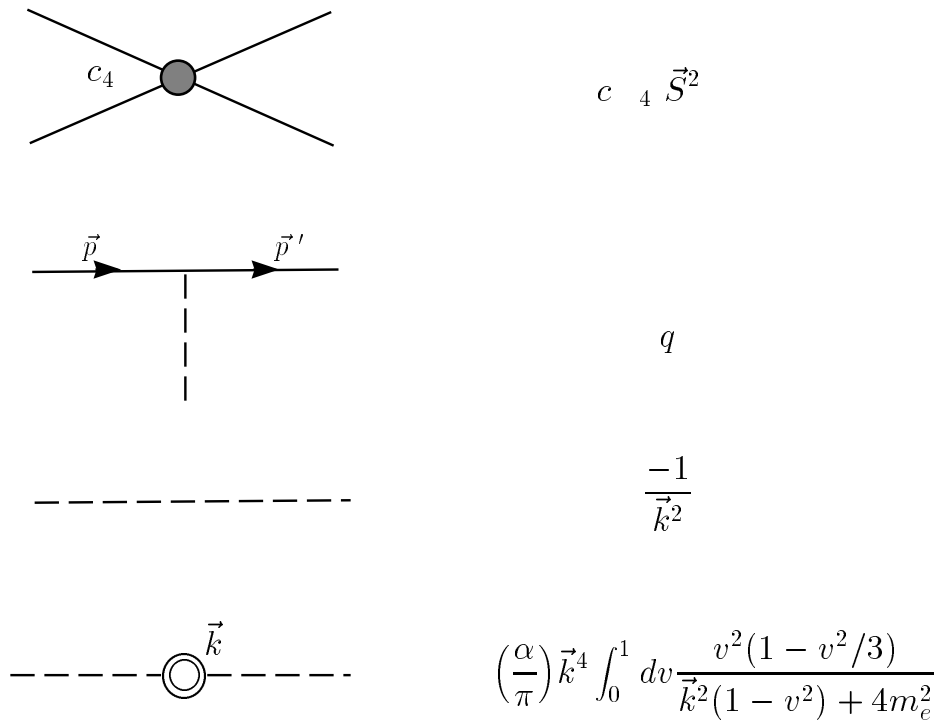}}
\caption{NRQED Feynman rules (q represents the charge of the particle).
The last three rules can also be applied to NRChPT.}
\label{pat}
\end{figure}
\begin{figure}
\centerline{\epsfxsize 5.0 truein
\epsfbox{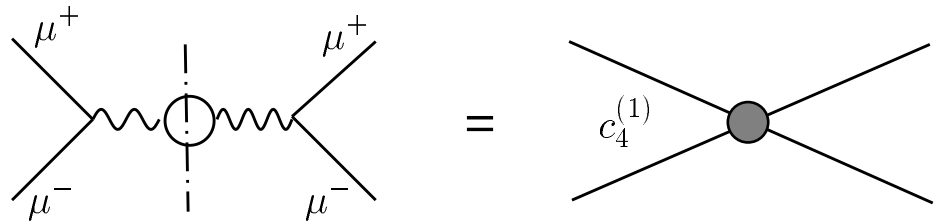}}
\caption{Lowest order matching of $
 {\rm Im} (c_4)$ (the imaginary part
is implicit in the figure).}
\label{pat3}
\end{figure}

\begin{figure}
\centerline{\epsfxsize 5.0 truein
\epsfbox{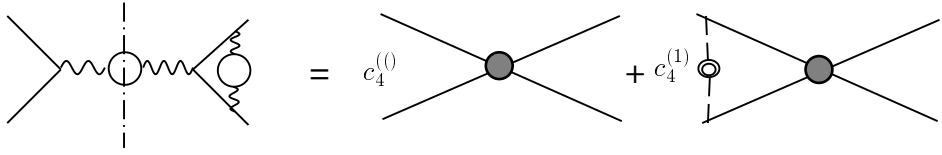}}
\caption{Vacuum polarization correction to
the coefficient ${\rm Im} (c_4)$ (the imaginary
part of the coefficients is implicit
in the figures).}
\label{pat1}
\end{figure}
\begin{figure}
\centerline{\epsfxsize 5.0 truein
\epsfbox{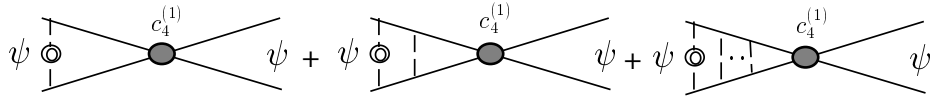}}
\caption{VPC correction to the dimuonium decay rate.
The three diagrams represent the no Coulomb, one Coulomb and R 
contributions (the imaginary part of the coefficients
is implicit in the figures).}
\label{pat2}
\end{figure}
\begin{figure}
\centerline{\epsfxsize 5.0 truein
\epsfbox{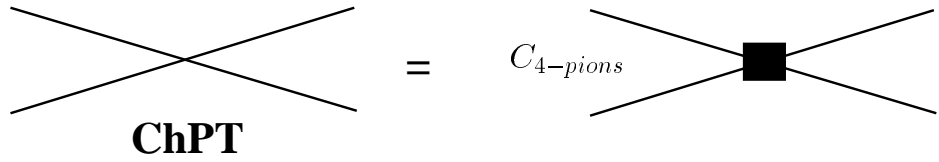}}
\caption{Lowest order matching of the ChPT and NRChPT
$\pi^+ \pi^- \pi^0 \pi^0$ interaction.}
\label{pat5}
\end{figure}
\end{document}